%% file: paper.tex
\documentclass[10pt,conference]{IEEEtran}

\IEEEoverridecommandlockouts

\ifCLASSINFOpdf
  \usepackage[pdftex]{graphicx}
  \graphicspath{{./images/}}
\else
  \usepackage[dvips]{graphicx}
  \graphicspath{{./images/}}
\fi
\usepackage{subfigure}
\usepackage[cmex10]{amsmath}
\usepackage{amssymb}
\usepackage{algorithm}
\usepackage{algorithmic}
\newtheorem{theorem}{Theorem}[section]

\newtheorem{corollary}{Corollary}[section]

\begin{document}
\title{Characterization of Random Linear Network Coding with Application to Broadcast Optimization in Intermittently Connected Networks}

\author{\IEEEauthorblockN{Gabriel Popa\IEEEauthorrefmark{1}
}

\IEEEauthorblockA{\IEEEauthorrefmark{1} Computer Engineering and Networks Laboratory, ETH Zurich, %
Gloriastr. 35, 8092 Zurich, Switzerland \\
Email: gpopa@tik.ee.ethz.ch}

\vspace{-1cm}
}

\maketitle %
\thispagestyle{empty}
\pagestyle{empty}
\thispagestyle{empty}
\pagestyle{empty}

\begin{abstract}
We address the problem of optimizing the throughput of network coded traffic in 
mobile networks operating in challenging environments where connectivity is intermittent 
and locally available memory space is limited. Random linear network coding (RLNC) is 
shown to be equivalent (across all possible initial conditions) to a random message 
selection strategy where nodes are able to exchange buffer occupancy information 
during contacts. This result creates the premises for a tractable analysis of RLNC packet 
spread, which is in turn used for enhancing its throughput under broadcast. By exploiting 
the similarity between channel coding and RLNC in intermittently connected networks, we 
show that quite surprisingly, network coding, when not used properly, is still significantly 
underutilizing network resources. We propose an enhanced forwarding protocol that 
increases considerably the throughput for practical cases, with negligible additional 
delay.

\end{abstract}

\input{introduction}
\input{network_models}
\input{packets_density_distribution}
\input{optimal_forwarding_protocol}

\input{simulation_results}

\input{conclusions}
\IEEEpeerreviewmaketitle

\bibliographystyle{IEEEtran}
\bibliography{IEEEabrv,paper}

\input{appendix-to-paper}

\end{document}

%% file: introduction.tex
\section{Introduction}
\thispagestyle{empty}

The paper focuses on improving throughput in intermittently connected networks while maintaining low delivery delays. 
Intermittently connected networks (or DTNs -- disruption tolerant networks) are networks of very mobile, power- and 
memory-constrained devices where connectivity is sporadic. This is the model of choice for wireless networks operating 
in challenging conditions (networks of UAVs, disaster relief scenarios, {\em etc.}). As traditional routing approaches 
cannot be applied in this case (little is known in advance about future connectivity), the literature has developed 
\emph{opportunistic (epidemic) forwarding} protocols that replicate packets to multiple relay nodes in order to optimize the 
delivery delay and/or the chance that packets get delivered to the destination(s)~\cite{Vahdat00epidemicrouting}. 
Increasing throughput, while keeping low delay, is a problem of practical interest as it would enable nodes to receive 
more information per time unit, with almost the same delay. %
RLNC has emerged recently as a promising approach for such applications. It ameliorates the transmission by 
introducing the diversity of multiple independent combinations in the epidemic forwarding. Nevertheless, analyzing 
and optimizing RLNC for DTNs is difficult\cite{WiOpt2010Fekri}. We prove that RLNC in DTNs is in fact equivalent 
to a forwarding algorithm not employing network coding, which is much easier to analyze. Using this equivalence, 
we show that RLNC still produces too many redundant packets during contacts, thereby underutilizing network resources 
(inter-contact times and consequently buffer space). Our study shows that transmissions of a backlogged source can 
be conveniently pipelined even though no feedback from destinations is available such that significant throughput 
gains can be attained with negligible additional delay. Based on these observations, we design and evaluate a 
forwarding protocol with reduced buffer and energy requirements (less mobility required for collecting the same 
number packets).

\emph{Related work:} In their seminal paper, Deb {\em et al.}\cite{Random_Network_Coding_Medard} offer 
an in-depth analysis of random linear network coding (RLNC) for networks with intermittent contacts. 
Numerous studies have built upon these results, extending them to the case of DTNs\cite{WidmerNetCod,BenLiangMobiOpp,BenLiangInfocom2008}; 
RLNC is used to improve the average delivery ratio within a given time unit. Our work is motivated by 
the observation that these studies extend the conclusions of \cite{Random_Network_Coding_Medard} past 
the assumptions under which those results have been obtained, such that they do not hold anymore. In 
particular, all protocols for optimizing throughput or delays in DTNs break the \emph{initial condition 
assumption} (messages do not have equal initial spread), which leads to significant throughput loss. 
Both Lin {\em et al.}\cite{BenLiangInfocom2008} and Altman {\em et al.}\cite{AltmanInfocomControl} (which 
studies network coding and Reed-Solomon codes in two-hop DTNs) note the similarity between channel 
coding and data transmission in DTNs. We study the implications of this analogy on the aforementioned 
initial condition assumption. 

%% file: network_models.tex
\section{Network Models}
\label{sec:buffer_network_coded_model}
\thispagestyle{empty}

The network model is similar to the one used in \cite{WiOpt2010Fekri}. The network consists 
of $N$ mobile nodes with the same radio range and buffer space $B$. We 
consider that a wireless link (contact) is established between two nodes when they are in 
each others' radio range. All contacts are bidirectional. Their duration is considered to 
be negligible with respect to the inter-contact times, but sufficient enough to allow the 
transmission of one packet in each direction. We consider mainly the case of a backlogged 
source broadcasting data to the entire network and then extend the conclusions to  
multi- and unicast. The source aims at maximizing the average throughput at destinations. 
We consider a mobility model with exponential inter-contact times of parameter $\lambda$, which 
has been validated for a wide spectrum of mobility scenarios\cite{GroeneveltSigmetrics2005}. 
Our analysis is however not constrained to this type of mobility. 
The backlogged source is considered to have at least $\nu \in \mathbb{N}$ packets in its buffer. 
These packets will be called hereafter \emph{variables} and represent original (not coded) 
source-generated packets. Out of them, the source selects every time a set of $\nu$ oldest 
packets to be transmitted in the network. When the transmission 
of the $\nu$ packets is considered completed (after a fixed time, TTL, set as a function of $N$), 
they are deleted by all nodes, the source selects the next $\nu$ packets 
and repeats the operation. Minimizing the time to delivery and the probability 
that not all nodes have decoded the content are desirable. 

All nodes implement random linear network coding over a finite field $GF({2^k})$. The source 
is assumed to send to nodes that it encounters coded packets (one such packet/contact). 
\emph{Coded packets} are elements in the set of $\nu$ independent linear combinations of 
$\nu$ variables (set called \emph{packet batch}), where coefficients are randomly selected from $GF({2^k})$. 
Note that 
$\nu \leq B$ and the buffer occupancy is described by the number of independent linear 
combinations present in a node's buffer. Packets have size $K$ (a multiple of $k$) and are 
treated as vectors of values from $GF({2^k})$. During a contact, nodes scale each 
vector (coded packet) in their buffers with randomly selected elements from $GF({2^k})$ and 
adds them, thereby creating a new network coded packet, which is sent to the other node. 
A node is able to decode all variables only when it has received $\nu$ independent linear 
combinations. We say that a coded packet received by a node is innovative if it increases 
the rank of the equation system formed by coded packets in that node's buffer. A contact 
is \emph{efficient} iff at least one innovative packet is transferred. We are analyzing 
two protocols: one in which relay nodes send random linear combinations of coded packets 
stored in their buffer during contacts (as described above) and the other where nodes 
compare their buffers\footnote{Using counting Bloom filters} and only forward to each 
other (coded) packets selected uniformly at random among those not contained by the other. The two protocols 
are denoted by $\Gamma$ (true RLNC) and $\Delta$ (a type of random message selection), 
respectively. RLNC schemes transport along with packets the random coefficients as well 
as the identities of original variables combined in the coded packets, providing therefore a distributed 
solution\cite{Ho2003Randomized,Chou2003Practical}. It can be proven that the overhead of 
storing and transporting these random coefficients is small. Note that $\Delta$ can also 
be used with variables as packets (instead of coded packets), as relays do not perform 
network coding, thus eliminating coefficient overhead. 
$\Gamma$ and $\Delta$ are similar to E-NCP, E-RP\cite{BenLiangInfocom2008}.

%% file: packets_density_distribution.tex
\section{Main Results}
\label{sec:packet_densities}
\thispagestyle{empty}

\subsection{Random Message Selection with Feedback {\em vs.} RLNC}

The following result shows that the operation of random message selection with buffer feedback 
during contacts ($\Delta$) is almost identical to true RLNC ($\Gamma$). Thus, results for  
$\Delta$ apply to $\Gamma$ and vice versa. The equivalence uncovered by this theorem can 
be used for designing optimal distributed network coding protocols for intermittently-connected 
networks, initially under the more tractable $\Delta$, then applied to $\Gamma$. $\Delta$ relies 
on nodes exchanging information about the list of packets in buffers, during contacts. Should 
this capability not be available, $\Gamma$-type RLNC offers the distributed counterpart. 

\begin{theorem} 
Given identical mobility and initial conditions (set of packets already disseminated by 
the source in the network and which are prepared to start the epidemic network spread), an 
arbitrarily-selected contact between two nodes $A$ and $B$ at time $t$ will have approximately the same 
probability that $A$ delivers a novel packet to $B$, under both $\Gamma$ and $\Delta$.\footnote{\cite{BenLiangMobiOpp} 
also implies that ($\Gamma \cong \Delta \cong $ global rarest).}
\end{theorem}

\begin{IEEEproof}
We use the following notation: for a node 
$w$, $S^{-}_w$ and $S^{+}_w$ designate the subspace spanned by the coded packets belonging to this node's 
buffer, before and after a contact with another node $v$, respectively. It is thus easy to 
infer\cite{Random_Network_Coding_Medard} that: 
\begin{eqnarray}
Pr[ S^{+}_w \nsubseteq S^{-}_u | S^{-}_w \subseteq S^{-}_u, S^{-}_v \nsubseteq S^{-}_u] & \geq & 1 - \frac{1}{q} \label{eqlemmed1}\\
Pr[ dim(S^{+}_w) > dim(S^{-}_w) | S^{-}_v \nsubseteq S^{-}_w ] & \geq & 1 - \frac{1}{q} \label{eqlemmed2}
\end{eqnarray}
where $q=|GF(2^k)|$ and $u$ can be any other node. The two probabilities describe the way a 
node $w$ acquires new degrees of freedom in its buffer under RLNC with $\Gamma$-type forwarding. Each 
of these probabilities is equal to $1$ for $\Delta$-type forwarding, so eq. \eqref{eqlemmed1} and \eqref{eqlemmed2} 
continue to be true even for $\Delta$. If we consider the case of a very large $q$ ($q \rightarrow +\infty$), 
then  $\Gamma$ and $\Delta$ become identical. For known mobility and known 
initial packet distribution, we can construct a DTMC to capture the packet propagation. A state contains 
an array of size $N$ and an element of this array at index $i$ has to store the list of degrees of 
freedom acquired by node $i$ until that time step. There are $N$ degrees of freedom under both $\Gamma$ and 
$\Delta$. Consider a contact between nodes $A$ and $B$, where we analyze only the transmission from $A$ to 
$B$. From eq. \eqref{eqlemmed1}, \eqref{eqlemmed2} this transmission is successful iff node $A$'s buffer 
has one degree of freedom not available to $B$. In both $\Gamma$ and $\Delta$ this degree of freedom is 
selected uniformly at random from those available to $A$ and not available to $B$. Thus, the transition 
probabilities are the same for both DTMCs and the two protocols behave identically. In a more realistic 
setting when $q$ is finite, RLNC with $\Gamma$ will in fact slightly underperform $\Delta$, because the 
probabilities in eq. \eqref{eqlemmed1}, \eqref{eqlemmed2} will be $1$ for $\Delta$ and $\geq 1 - \frac{1}{q}$ 
for $\Gamma$.

To prove rigorously that the uniform selection of degrees of freedom (dimensions) leads to similar 
behavior of $\Gamma$ and $\Delta$, we have to postulate the following elementary theorem, known from 
linear algebra, presented here without proof: 

\textbf{Every $n$-dimensional vector space $V$ over some finite field $F$ is isomorphic to $F^n$. If $v_1, v_2, v_3, \ldots, v_n$ is 
a basis of $V$, then the mapping $\phi: F^n \rightarrow V$ : $(a_1, \ldots, a_n) \mapsto \sum_{k=1}^n a_k v_k$ is an 
isomorphism. }

\emph{Observation:} Since the choice of basis for V is not unique (there are many possibilities) $\Rightarrow$ 
the above isomorphism is also not unique. In fact, we can construct many such isomorphisms. 

\emph{Final steps:} We need this isomorphism simply because tracking the evolution of vector spaces (that is, 
node buffers) during the packet spread process is very challenging. Such isomorphisms offer an easy way to 
\emph{label buffers} in a consistent manner. In particular, we are interested in mapping each buffer to a 
\emph{subset of the base} $\{p_1, \ldots, p_{\nu}\}$, where $p_1, \ldots, p_{\nu}$ are the initial packets at 
source. We regard each buffer as a subspace/subset of the $\nu$-dimensional vector space. Each such buffer/subspace 
is generated by the vectors/packets present in it. Note that the labelling will be performed for every node, 
at will hold at every step of the packet spread. However, a final point needs to be discussed. One has to 
observe that we cannot simply map all $k$-dimensional subspaces to the same set of $k$ vectors of the base (actually, 
to the subspace that they generate). This is simply because then all buffers will look identical after applying 
the isomorphism. Based on the fact that every intersection of subspaces is also a subspace, we can build the 
mappings/labellings for each node in a way that can prevent this problem. To this end, we specify a hard 
constraint \emph{requiring that the intersection of subspaces be respected even after applying the isomorphism}. 
This can be translate as follows: the intersection of any number of subspaces (buffers) has to be a subspace of the 
same dimension in the original version \emph{and} after applying the isomorphism. This is effectively the final 
step of our proof. The attentive reader will have already noticed that instead of working with coded packets, we 
have mapped our buffers to sets of original packets, thus effectively equating $\Gamma$ to $\Delta$. 
\end{IEEEproof}

\subsection{Finding Optimal Spread Ratios}
\label{sec:optimal_densities}
We analyze how the number of coded packet copies influences the instant throughput of broadcast 
under $\Delta$-type forwarding and extend the result to $\Gamma$ forwarding, uni- and multicast. 
For our mobility model, if coded packets have each the same number of copies in the network 
at the beginning of the forwarding, then, for an arbitrary $k$, $Pr[$node $\ell$ has a copy 
of coded packet $p_k]=\zeta_k=ct., (\forall) \ell$. $m_{p_i}(t)$ is the number of copies of 
$p_i$ contained by network at time $t$ (not counting the source), and $\rho_{p_i}(t)=\frac{m_{p_i}(t)}{N-1}$ 
is the correspondent instant density. %
We seek to find the relation between $\rho_{p_i}(t), i=\overline{1, \nu}$ that maximizes the 
instantaneous throughput. 
For this, we analyze the efficiency of each node's first contact after instant $t$, 
arbitrarily chosen. For tractability, we first look at the case when 
\emph{each relay node contains exactly one coded packet} at time $t$
and generalize %
afterwards. In this case, 
$\sum \limits_{i=1}^{\nu} m_{p_i}(t)=N-1, \sum \limits_{i=1}^{\nu} \rho_{p_i}(t)=1$. 
We exclude \emph{w.l.o.g.} the source as its contacts are efficient \emph{by definition} 
anyway. 
If ${\cal A}_{p_i}^{(t)}$ is 
the set of nodes (without the source) containing a copy of coded packet $p_i$ at time $t$, 
$|{\cal A}_{p_i}^{(t)}|=m_{p_i}(t)$, $(\forall) i=\overline{1, \nu} \Rightarrow \bigcup \limits_{i=1}^{\nu} {\cal A}_{p_i}^{(t)}={\cal N} - \{s\}, {\cal A}_{p_i}^{(t)} \cap {\cal A}_{p_j}^{(t)} = \emptyset, (\forall) i \neq j$, 
where ${\cal N} - \{s\}$ is the set of all nodes, without the source. For an arbitrary 
$\ell \in {\cal A}_{p_i}^{(t)}, Pr[$next contact of $\ell$ is inefficient$]=\rho_{p_i}(t)-\frac{1}{N-1}$ 
(efficient with probability $1-\rho_{p_i}(t)+\frac{1}{N-1}$), meaning that $\ell$ has met another node from the 
same set, no data transfer occurred and the waiting time preceding the contact had been wasted. 

We are interested in maximizing throughput (maximizing the expected number 
of efficient first contacts of each node after instant $t$). Therefore, 
under $\sum \limits_{j=1}^{\nu} \rho_{p_j}(t) = 1$ we maximize\footnote{Considers bidirectional contacts. Approximation that node densities do not change significantly between two contacts verified in Section \ref{sec:impact_of_densities}. The constant $\frac{1}{N-1}$ does not influence the result.}
\begin{eqnarray}
f(\rho_{p_1}(t), \ldots, \rho_{p_{\nu}}(t)) & = & \sum \limits_{k=1}^{\nu} \sum \limits_{\ell \in {\cal A}_k^{(t)}} (1-\rho_{p_k}(t)) = \\
= \sum \limits_{k=1}^{\nu} m_{p_k} (1-\rho_{p_k}) & = & (N-1) \cdot \sum \limits_{k=1}^{\nu} \rho_{p_k} (1-\rho_{p_k})
\end{eqnarray}
Using Lagrange multipliers, 
\begin{eqnarray}
\Lambda (\rho_{p_1}(t), \ldots, \rho_{p_{\nu}}(t), \lambda_{sol})  =  f ( \rho_{p_1}(t), \ldots, \rho_{p_{\nu}}(t))+ \\
\lambda_{sol} \cdot (\sum \limits_{k=1}^{\nu} \rho_{p_k}(t) - 1)  \Rightarrow  \frac{\partial \Lambda}{\partial \rho_{p_k}} = (N-1) \cdot (1-2\rho_{p_k}) \\
+\lambda_{sol} = 0 \Rightarrow \lambda_{sol}=\frac{2(N-1)}{\nu}-(N-1) %
\end{eqnarray}
Replacing $\lambda_{sol}$ we find that 
$\rho_{p_1}(t)=\ldots=\rho_{p_{\nu}}(t) \triangleq \rho(t) = \frac{1}{\nu}$. 
Thus, all densities should be equal at instant $t$. The generalization for the case of more 
(or less) than $1$ packet/node on average ($\sum \limits_{j=1}^{\nu} \rho_{p_j}(t) = c, c \geq 0$) is 
provided by 
\begin{theorem}
The following condition is necessary for maximizing throughput of a batch of $\nu$ 
coded packets in a DTN with $\Delta$-type forwarding:
$\rho_{p_1}(t)=\rho_{p_2}(t)=\rho_{p_3}(t)=\ldots=\rho_{p_{\nu}}(t) \triangleq \rho(t), (\forall) t$. 
In other words, regardless of the buffer occupancy level, packet densities 
should be roughly equal to ensure maximal throughput. 
\end{theorem}
\begin{IEEEproof}
For each node $\ell$ define the concept of \emph{entire buffer packet} as being the indicator function 
$1_{\ell}: \{p_1, p_2, p_3, \ldots, p_{\nu}\} \rightarrow \{0,1\}, 1_{\ell}(p_k) = 1 \Leftrightarrow$ 
the buffer of $\ell$ contains coded packet $p_k$. A contact between $\ell, \aleph \in {\cal N}$ is efficient 
$\Leftrightarrow 1_{\ell} \neq 1_{\aleph}$. Using the above argument for \emph{entire buffer packets} we check 
that $\sum \limits_{{\ell} \in {\cal N}-\{s\}} \rho_{1_{\ell}} = 1$ and therefore 
$\rho_{1_{\ell}}(t)=\rho_{buf}(t)=ct., (\forall) \ell \in {\cal N}-\{s\}$ is necessary for throughput maximization. 
From this set of equalities at time $t$, considering an arbitrarily chosen but fixed \emph{entire buffer packet} 
$1_{\aleph} \Rightarrow Pr[\ell \in {\cal A}_{1_{\aleph}}^{(t)}] = Pr[1_{\ell}(p_1)=1_{\aleph}(p_1)] \cdot Pr[1_{\ell}(p_2)=1_{\aleph}(p_2)] \cdot Pr[1_{\ell}(p_1)=1_{\aleph}(p_1)] \cdot \ldots \cdot Pr[1_{\ell}(p_{\nu})=1_{\aleph}(p_{\nu})]=\rho_{buf}(t)=ct., (\forall) \ell \in {\cal N}-\{s\}$.\footnote{The independence assumption is reasonable, given that a coded packet is selected by a node randomly from its buffer for transmission during a contact (only from those packets that are \emph{innovative}).} 
But since none of the buffers is yet full ($\max_{{\ell} \in {\cal N}-\{s\}} |1_{\ell}|=\kappa < N$, $|1_{\ell}|=\kappa_{\ell}$ 
-- $\ell$'s buffer occupancy) $\Rightarrow$ 
$Pr[1_{\ell}(p_{i_1}) = 1] \cdot Pr[1_{\ell}(p_{i_2}) = 1] \cdot Pr[1_{\ell}(p_{i_3}) = 1] \cdot \ldots \cdot Pr[1_{\ell}(p_{i_{\kappa_{\ell}}}) = 1]=\Omega=ct., i_1, i_2, i_3, ..., i_{\kappa_{\ell}}=\overline{1, \nu}, (\forall) \ell \in {\cal N}-\{s\}$. 
This is an equation system with $N-1$ product equations and $\nu \leq B \leq N$ unknowns, where each probability is known 
to be strictly positive. By equalizing equations with the same number of factors, we obtain that with high probability 
$Pr[1_{\ell}(p_1) = 1] = Pr[1_{\ell}(p_2) = 1] = Pr[1_{\ell}(p_3) = 1] = \ldots = Pr[1_{\ell}(p_{\nu}) = 1]=ct., (\forall) \ell \in {\cal N} - \{s\}$, which means that coded packets should have equal spreads. 
\end{IEEEproof}

\emph{Remark:} The same condition is necessary for providing optimal throughput also for unicast and multicast. This 
is because each node will deliver a packet to the target(s) with equal probability. %

\subsection{Impact of Packet Counts on Contact Efficiency}
\label{sec:impact_of_densities}

\begin{figure}[t!]
  \centering
   \subfigure[Protocol 1 (continuous source transmission) {\em vs.} Protocol 2. (equal densities)]
    {
      \includegraphics[scale=0.55]{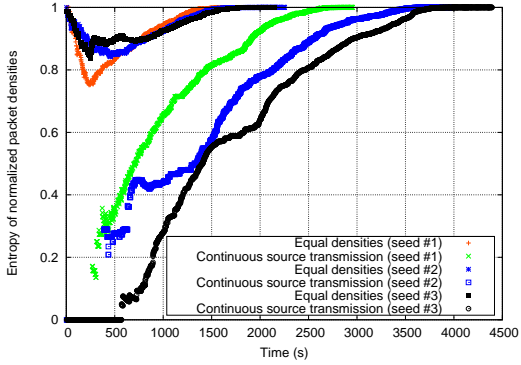}%
      \label{fig:11_11_nodes_sources_help}
    } %
    \subfigure[Protocol 1 (continuous source transmission) {\em vs.} Protocol 3. (equal densities)]
    {
      \includegraphics[scale=0.55]{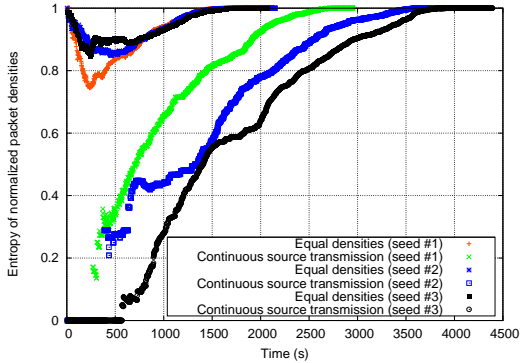} %
      \label{fig:11_10_nodes_source_does_not_help_equal_densities}
    }
   \vspace{-0.15cm}
   \label{fig:entropies}
   \caption[Optional caption for list of figures]{Comparison of instantaneous entropies of normalized network coded packet densities for a network of $100$ nodes}
   \vspace{-0.8cm}
\end{figure}

In this paragraph we explain why the assumption of equal packet spread cannot be taken for granted and 
demonstrate its performance impact. %
We define the entropy of relative (normalized) coded packet densities 
at time instant $t$ as $H(\rho'(t)) = \sum \limits_{k=1}^{\nu} \rho'_{p_k}(t) \cdot \log _{\nu} (\frac{1}{\rho'_{p_k}(t)})$, 
where $\rho'(t)=(\rho'_{p_1}(t), \rho'_{p_2}(t), \rho'_{p_3}(t), \ldots, \rho'_{p_{\nu}}(t))$ and $\rho'_{p_k}(t) \in [0,1]$ \footnote{We consider by convention that $0 \cdot \log(+\infty) = 0$.}are 
the normalized counterparts of $\rho_{p_k}(t), (\forall) k=\overline{1, \nu}$. The entropy is close to $1$ 
\emph{iff} network coded packets have similar instant densities in the network. %
The entropy allows us to quantify the discrepancy between densities through a scalar. 
We analyze the evolution with time of $H(\rho'(t))$ %
and use as an example a network with $N=100$ nodes, $B=11$, $\lambda = 0.005$. \footnote{The same behavior is observed for other combinations of parameters. Timeline scaled with $N-2$ in plots.} 
In Fig. \ref{fig:11_11_nodes_sources_help}-\ref{fig:11_10_nodes_source_does_not_help_equal_densities}
 we show how the entropy evolves over time using three representative mobility realizations that run until all destinations decode the data, 
for the following protocols: 
\begin{itemize}
\item[1.] A benchmark $\Delta$-type protocol: %
the source transmits a batch of $11$ coded packets continuously until all destination nodes have decoded the data. No specific measure is taken to maintain equal densities; %
\item[2.] Another $\Delta$-type protocol with a batch of $10$ coded packets, each placed in a separate node before the spread is triggered. The source continues transmitting coded packets to non-full nodes; 
\item[3.] Similar to 2., with the exception that after distributing the initial copies and triggering their spread, the source stops disseminating data. \footnote{Fig. \ref{fig:11_11_nodes_sources_help},\ref{fig:11_10_nodes_source_does_not_help_equal_densities} show time-evolving entropies for protocols 2., 3. (equal densities) without the time needed by the source to place a copy of each coded packet in disjoint nodes. This will be studied in Section \ref{sec:optimal_forwarding_protocol}.}
\end{itemize}

\emph{Remarks:} There are a number of observations, which hold in general (also for $\Gamma$). Firstly, high entropies are 
conserved by exponential inter-contact times. Secondly, the delay of protocols 2. and 3. is always almost 
identical, meaning that the source intervention does not improve the throughput anymore and that high 
entropy should be sufficient for maximizing throughput. Thirdly, when the source injects packets in a greedy 
manner (strategy commonly considered to yield minimal delay\cite{BenLiangInfocom2008}), the entropy drops significantly, 
impacting overall contact efficiency. 

%% file: optimal_forwarding_protocol.tex
\section{Improved Forwarding Protocol}
\label{sec:optimal_forwarding_protocol}
\thispagestyle{empty}

We define the \emph{seeding phase} of a transmission as the time interval used by source to place $\nu$ 
independent coded packets each on a distinct relay node, for a batch of $\nu$ variables. The 
time needed for this operation is a random variable $T^s_n$, where $n \in \mathbb{N}$ is the identifier 
of the packet batch (or $T^s$ when we do not refer to a specific $n$). Similarly, we define the 
\emph{propagation phase} of the transmission as the interval in which the independent coded packets 
are forwarded epidemically in the network; this step finishes when all destination nodes have successfully 
decoded the packets. The time needed for the propagation phase of batch $n$ is another random variable, 
$T^p_n$ (identically distributed as $T^p$). For every packet batch, the propagation phase takes place 
immediately after the seeding phase. The key idea is that the seeding phase of packet batch $n+1$ can be performed 
in parallel with the propagation phase of the packet batch $n$. To accomplish this, we need to ensure that
$\nu=B-O({\cal C})$, s.t. $B-O({\cal C})$ buffer slots are available to propagation of batch $n$ and 
$O({\cal C})$ are reserved for seeding of batch $n+1$. For an each node, these $B-\nu$ places will host 
packets copied directly from the source. 

\emph{Remark:} The throughput loss caused by the fact that $\nu \neq B$ is shown to be negligible in 
comparison to the gain resulting from pipelining (see Section \ref{sec:simulation}). In practice 
$B-\nu \in \{1, 2\}$. 

\begin{theorem}
The seeding phase can be completed in $\Theta(\nu)$ steps (in practice, approximately $\nu$ consecutive contacts of the source) with 
high probability. At the end of the seeding phase, each of the $\nu$ independent coded packets 
will be placed on a different relay node with high probability.
\end{theorem}
\begin{IEEEproof}
The \emph{seeding algorithm} performed by the source is described in the following. From the original $\nu$ 
variables, the source constructs $\nu$ \emph{independent} coded packets with RLNC. Each of these $\nu$ coded 
packets is sent by the source only once. During $j^{th}$ contact, the $j^{th}$ coded packet $p_j$ is 
sent by source to the peer node, $j = \overline{1, \nu}$. To ensure that all packets start spreading 
at roughly the same time (during propagation phase), the source specifies that coded packet $p_j$ 
should be forwarded only after the estimated time to finish the seeding. 
In the most favorable case, the source encounters a different node every time. This happens for 
$B \ll N$ ($\nu \ll N$). In this case we can set $\nu = B-1$. Let $P'_i$ be the probability 
that packet $p_i$ will be successfully placed in a node not already containing a packet from the 
same batch. Then, $P'_i=\frac{N-i+1}{N} \Rightarrow E[X]=\sum \limits_{i=1}^{\nu} \frac{1}{P'_i}$, 
where $X$ is a r.v. (the number of steps to perform seeding). Thus $E[X] \approx \nu$ for $B \ll N$. The source 
faces a variant of the coupon collector problem for higher $B$. For this case, we can set 
$\nu = B-2$, and the probability that two coded packets of the same batch end up in the same node during seeding is much higher. 
However, the relay will move the extra coded packet to another node not already containing a coded 
packet from the same batch) with the first opportunity, 
which occurs with high probability. Therefore, the probability that $p_i$ is not placed successfully 
for this case (neither by the source, nor by the relay at some point in the future, before the end of 
the seeding phase, which should occur after the $\nu^{th}$ contact of the source) is 
$\pi_i \leq (1-P'_i) \cdot \prod \limits_{k=i+1}^{\nu} (1-P'_k)=\frac{i-1}{N} \cdot \prod \limits_{k=i+1}^{\nu} \frac{k-1}{N}$. 
In practice, we work with networks of limited buffers, were $\nu \leq B \ll N$ and therefore $\pi_i \approx 0$. 
This probability is very low even when $\nu=\lceil \frac{N}{2} \rceil$. %
In conclusion, seeding can be done on average in $\nu$ steps successfully. 
\end{IEEEproof}
Splitting in two phases (seeding and propagation) is suggested by the resemblance to channel coding: to 
approach channel (which is analogous to the DTN) capacity, a block of $\nu$ bits is assembled, coded, sent 
and then decoded by the destination. As $\nu \rightarrow +\infty$, the capacity can be approached 
asymptotically. 

\begin{corollary}
The seeding phase occurs with the minimum possible energy consumption for the source. 
\end{corollary}

No feedback is assumed and reliable packet delivery is required even when the source is backlogged. 
We therefore enforce $T^l_p$ as deadline for the propagation phase and aim to achieve full delivery 
with high probability, before this deadline is reached. The time spent in the propagation 
phase is measured from the $\nu^{th}$ contact of the source (the one that delivered the last packet of 
the batch to the network). The probability that the propagation phase will be longer than $T^l_p$ can 
be obtained using one of the following: 
\begin{itemize}
\item $Pr[T^p > T^l_p] \leq \frac{E[T^p]}{T^l_p} = \varepsilon_p$ (Markov's inequality);
\item $Pr[T^p > T^l_p] \leq inf_s e^{-sT^l_p} \cdot M_{T^p}(s) = \varepsilon_p$, where $M_X(s)$ is the\
mgf of variable $X$ (Chernoff's inequality which is the tightest, if applicable).
\end{itemize}
The derivation for $T^p$ (CCDF) is omitted here due to limited space and is provided by \cite{AkisStoppingTimes}. 
Moreover, due to the fact that packet densities are almost equal at the beginning of the propagation phase, 
the assumptions made in \cite{BenLiangInfocom2008} 
are now accurate, allowing easier analytical treatment. The probability that there is at least one destination 
that has not decoded all data is $\varepsilon_p \rightarrow 0$, for $T^l_p$ reasonably large. In Section 
\ref{sec:simulation} we show that this condition can be achieved already for low $T^l_p$ and therefore 
throughput is not affected.

%% file: simulation_results.tex
\section{Simulation Results}
\label{sec:simulation}
\thispagestyle{empty}

\begin{figure}[t!]
  \centering
   \subfigure[Throughputs ($\lambda=0.005,N=100$)]
    {
      \includegraphics[scale=0.28]{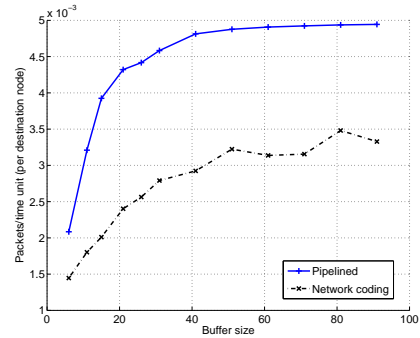}
      \label{fig:throughput_comparison}
    } %
    \subfigure[Delays ($\lambda=0.005,N=100$)]
    {
      \includegraphics[scale=0.28]{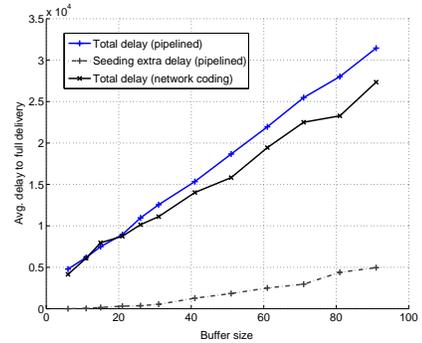}
      \label{fig:delay_comparison}
    }
   \vspace{-0.15cm}
   \label{fig:performance_comparison}
   \caption[Optional caption for list of figures]{Performance evaluation of the pipelining protocol (averaged over 100 source-generated packet batches)}
   \vspace{-0.8cm}
\end{figure}

We test the pipelined-$\Gamma$ protocol (with RLNC at intermediary nodes) against the simple 
$\Gamma$ forwarding protocol, which also uses coding at intermediary nodes and which should be 
throughput optimal. Fig. \ref{fig:throughput_comparison} shows the additional throughput provided 
by the pipelined protocol. To ensure full delivery, we let $T^l_p=\max_{n \in \{1, \ldots, 100\}} \{T^p_n\}$. 
Surprisingly, pipelining is quite close to achieving the 
throughput capacity (no more that one packet, coded or not, can be sent by the source during a 
contact). Fig. \ref{fig:delay_comparison} shows an extra delay incurred by packets due to 
the seeding phase exceeding the length of the propagation phase; its impact is however minimal. 
Using smaller buffers, pipelining can achieve throughputs superior to usual RLNC schemes (which need 
more memory), at the cost of a small additional delay. The pipelining protocol can be used also 
with non-coded packets. This is necessary when destination nodes only require some of the packets to be 
delivered, and do not need to decode the entire packet batch. In this case the overhead associated 
with transmission of coefficients and computations over the finite field is eliminated, but the 
observations from Fig. \ref{fig:throughput_comparison} and \ref{fig:delay_comparison} remain valid. 
The attempt made in \cite{BenLiangInfocom2008} to use equalizing spray counts does not obtain better delays 
simply because it still allows a long initial low entropy interval, which has a snowball effect. Further 
increasing the throughput by setting $\nu > B$ is not possible, because in broadcast every node must 
be able to decode the transmission. 

Our conclusions would seem to contradict the results of Ahlswede {\em et al.}\cite{Network_Information_Flow} 
and Deb {\em et al.}\cite{Random_Network_Coding_Medard}; this is however not the case, because  
we complement in fact the two papers for the case of intermittently connected networks. Firstly, 
RLNC would reach the maxflow-mincut bound when $\nu \rightarrow + \infty$, which would mean very large 
buffers (not possible in our case). Secondly, \cite{Random_Network_Coding_Medard} assumes equal initial 
packet spread, assumption which does not hold in DTNs with usual forwarding protocols.

%% file: conclusions.tex
\section{Conclusions}
\thispagestyle{empty}

In this paper we consider the problem of optimizing throughput in intermittently connected networks, with minimal impact on delay.
We specifically address the practical case of limited buffers. %
It is proven that network coding underutilizes the available resources. Following 
information theoretical hints, we design a practical forwarding protocol relying on pipelining, which achieves 
asymptotically reliable delivery and outperforms network coding in throughput, energy consumption and 
memory usage with negligible delay overhead. DTNs are shown to be very sensitive to initial forwarding conditions 
(in particular, initial number of packet copies). Setting them to convenient values is easily achieved and 
yields significant performance gains. On the other hand, trying to control the network after the initiation of 
the forwarding process is much more challenging. Our analysis of the single source broadcast generalizes to uni- and multicast. 
A thorough consideration of energy constraints, congestion, multiple unsynchronized sources, comparison with other 
coding techniques, improved pipelining and applicability of the maximum-entropy principle to other mobility models 
is left for future work.   

%% file: appendix-to-paper.tex
\section*{Appendix}
\subsection{Remarks Regarding the Equivalence Between $\Delta$ and $\Gamma$}
To see why the equivalence holds, we can consider the effect of the initial packet density distribution 
on both schemes. Let us assume that out of the $\nu$ coded packets, the source has managed to send to the network 
only $\psi < \nu$. These packets have the normalized density distributions $\rho'_1, \rho'_2, \rho'_3, \ldots, \rho'_{\psi}$, 
meaning that $\rho'_j=0, (\forall) j=\overline{\psi+1,\nu}$. Clearly, due to the uniformity of the mobility model, 
the probability distribution for having already received these packets is the same across nodes. Let us assume 
\emph{w.l.o.g.} that no coded packets from the source have been yet coded together. The higher the entropy, the 
higher is the chance that the distribution is very skewed. This means that some packets have already achieved high 
spread, while most others have only a few copies in the network. Then, the chance that two nodes in the next contact 
have exactly the same buffer content is very high. In this case, both $\Gamma$ and $\Delta$ generate the same 
inefficient contacts. Even if their buffers are not exactly the same, the overlap will be anyway significant. 
The contact will deliver with a high probability an independent packet to destination, but the problem is that 
most contacts in the network generate new random vectors \emph{from the same very few linear subspaces of similar 
dimension}. For this reason, a node having received a network coded packet is still very likely to deliver during 
its next contact a vector which is \emph{already in the linear subspace of the receiving node}. The essential 
observation to be made is that $\Gamma$ does not promote packets of lower densities better than $\Delta$. A rare 
packet will be coded together with others at basically the same rate as the one at which $\Delta$ promotes it. Indeed, 
the nodes receiving a rare packet will be able to deliver new combinations to others (under $\Gamma$), since the 
combinations contain the new packet. But this happens exactly the same under $\Delta$ too, anyway. As the buffers 
will be almost identical, the nodes having the rare packets will have to send them anyway, just like in $\Gamma$, 
because almost all the others that they have are already present in the nodes they meet. In other words, nodes 
receive new degrees of freedom at the same rate, both under $\Delta$ and $\Gamma$. What matters, is that a new 
independent vector has been received, but also the way it was obtained. If most nodes receive independent vectors 
generated from the same few bases, then in the next step they will for sure deliver redundant packets. As the 
source disseminates the \emph{initial base} $\{p_1, p_2, p_3, \ldots, p_{\nu}\}$ in the network during its contacts, 
it matters which packets of the initial base have reached destination nodes, and not the way these packets have 
been combined by the network. The assumption we made above that we first regard a network which has not coded 
yet packets together is indeed without loss of generality precisely for this reason. These simple facts provide 
us with the result that the behavior of both $\Delta$ and $\Gamma$ is almost identical. $\Delta$ can therefore 
be used as a very good approximation for $\Gamma$, where this is necessary for tractability reasons. 

\newpage
\subsection{Impact of Entropy on Contact Efficiency}

\begin{figure}[h]
   \subfigure[Random mobility with seed no. 1]
    {
      \includegraphics[scale=0.55]{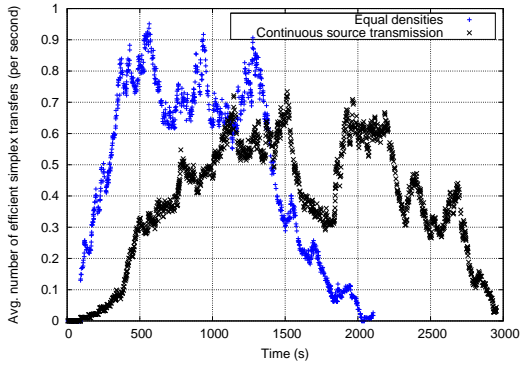} %
      \label{fig:speeds_for_randseed_1}
    }%
    \vspace{0.5cm}
    \subfigure[Random mobility with seed no. 2]
    {
      \includegraphics[scale=0.55]{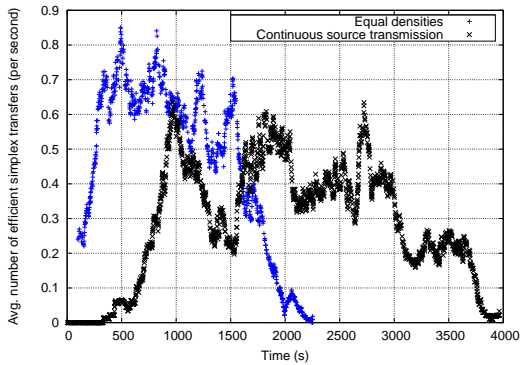} %
      \label{fig:speeds_for_randseed_2}
    }%
    \vspace{0.5cm}
    \subfigure[Random mobility with seed no. 3]
    {
      \includegraphics[scale=0.55]{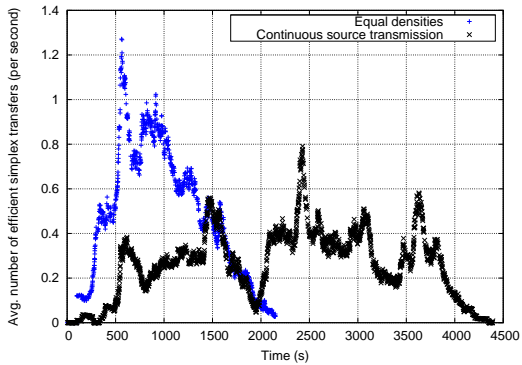} %
      \label{fig:speeds_for_randseed_3}
    }
   \vspace{-0.15cm}
   \label{fig:speeds}
   \caption[Optional caption for list of figures]{Average number of efficient contacts per unit of time (over a sliding window of 50 contacts), for protocols 1. (black) and 3. (blue). }
   \vspace{-0.8cm}
\end{figure}

Fig. \ref{fig:speeds_for_randseed_1}-\ref{fig:speeds_for_randseed_3} show contact efficiency for the same 
three mobility traces used in Fig. \ref{fig:11_11_nodes_sources_help}-\ref{fig:11_10_nodes_source_does_not_help_equal_densities}\footnote{Bidirectional contacts where both 
nodes have novel information for the other are counted as two efficient contacts.}. It can be clearly 
seen that high entropies allow the number of efficient contacts per unit of time to increase very fast 
and to remain at high levels, therefore improving throughputs, as opposed to the low entropy case. Low 
entropy will always generate much less efficient contacts, with negative effects as both ways of the 
bidirectional links established during contacts are affected.

